# DesignCon 2010

# PWB Manufacturing Variability Effects on High Speed SerDes Links: Statistical Insights from Thousands of 4-Port S-Parameter Measurements

Bart O. McCoy, Mayo Clinic
mccoy.bart@mayo.edu

Robert W. Techentin, Mayo Clinic

Benjamin R. Buhrow, Mayo Clinic
buhrow.benjamin@mayo.edu

Kevin Buchs, Mayo Clinic

How Lin, Endicott Interconnect

Barry K. Gilbert, Mayo Clinic
gilbert.barry@mayo.edu, 507-284-4056

Erik S. Daniel, Mayo Clinic

## Abstract
Variability analysis is important in successfully deploying multi-gigabit backplane printed wiring boards (PWBs) with growing numbers of high-speed SerDes links. We discuss the need for large sample sizes to obtain accurate variability estimates of SI metrics (eye height, phase skew, etc).

Using a dataset of 11,961 S-parameters, we demonstrate statistical techniques to extract accurate estimates of PWB SI performance variations. We cite numerical examples illustrating how these variations may contribute to underestimated or overestimated design criteria, causing unnecessary design expense. Tabular summaries of performance variation and key findings of broad interest to the general SI community are highlighted.

## Introduction
As data rates continue to rise to higher multi-gigabit per second speeds, it has become increasingly important to understand the performance of PWB transmission structures in backplane applications. Link performance margin is often tight, even when pre-emphasis and various types of equalization are employed.

The practical economics of business seeks maximum performance from the lowest cost materials, creating a tension between low-cost and the need for speed. The cost-performance curve for more exotic, lower loss PWB materials is steep. For example, moving a large PWB design into low-loss PTFE-based dielectrics adds extra cost and presents manufacturing problems such as high-speed drilling. The use of lower cost materials can be prolonged using techniques such as via back-drilling, but at significant cost and effort. While few can afford the price of over-engineering even fewer can afford the cost of systems failing in the field. These trends have bred a demand for a more detailed understanding of PWB manufacturing process variations and how these variations ultimately appear as electrical performance variations [15]. A more accurate, empirically driven understanding of PWB variability would clarify the performance boundary conditions, allowing engineers to extract maximum performance from lower cost materials with confidence.

Empirical information regarding PWB manufacturing variability (let alone the associated electrical variability) is difficult to come by, and may even be regarded as confidential by some fabricators. Generally, the only readily available information is that obtained by post-fabrication tests such as opens, shorts, capacitance and TDR testing. However, these tests do not fully characterize the high-speed PWB transmission line performance and they do not assess the PWB in the broader context of high speed link simulation. In this paper, we address both issues.

The impact of manufacturing variability on performance is regularly simulated in integrated circuit (IC) design using models specific to different IC process corners. Physical parameters of models are thoroughly characterized and variability can be studied as shown in [17,18]. In contrast, full SerDes link analyses often do not include the effects of PWB variability. In principle, PWB models could be characterized similar to models the IC industry. In this paper, we briefly address the limitations extracting performance variability from manufacturing variability data.

Instead, we focus on direct measurement and analysis of SI variables, utilizing recently developed tools [2-3] to measure thousands of 4-port S-parameters from several large PWBs and evaluate the PWB performance in the context of full link simulations. We illustrate, by example, how a direct PWB performance variability analysis may be performed and used with great benefit. S-parameter measurements transform all the important manufacturing *process variations* into signal integrity *performance variations*. In addition to raw S-parameter analysis, secondary SI quantities were computed including random skew, impedance and eye diagram metrics from full link simulations. In particular, we examine the statistical impacts and potential causes of phase skew as previously discussed in [4, 10-14]

We present a statistical analysis of this data, quantifying global variation and same-net variation. A multivariable ANOVA analysis reveals which variables were most important in predicting electrical performance variation. We show that a direct approach to electrical performance variability analysis tells the story of variation in the language which electrical engineers can quantify, specify and understand most readily.

## Paper Outline

We begin the paper with a brief discussion of other approaches to PWB variability analysis, highlighting our approach which emphasizes large numbers of direct measurements of SI variables of interest. We then describe the physical attributes of the PWBs we analyzed, the database of measurements collected and the various SI outcomes of interest such as impedance, eye height, phase skew, $S_{DD21}$, etc.

Next, we discuss to the statistical analysis techniques used in this paper (e.g. ANOVA, blocking, summary statistics, etc.) and provide a brief tutorial on how these techniques work, how they are useful, how to interpret their results and why they are necessary to properly analyze PWB variation. We also discuss the impact of non-normal distributions on the validity of the results.

The remainder of the paper is devoted to analyzing the data. We start with tables of summary statistics including global variability (all data lumped together) and same-net variability from pooled variance, an aggregate measure of variability of the same net across multiple boards. Finally, we provide extended discussions on four SI outcomes: s-parameters, phase skew, impedance and eye diagrams. We conclude with a discussion of the applications of PWB variability analysis.

## SI Variability from PWB Manufacturing Data

In principle, one may use physically parameterized models or equations to relate manufacturing process variables (such as line width and permittivity) to high speed signal integrity performance metrics (such as S-parameters). Corner case variation in manufacturing variables may be used to calculate expected variation in SI variables such as line impedance, as shown in [1].

Initially, we pursued this line of investigation with the help of the PWB fabricator. Detailed measurements of parameters such as line width, dielectric thickness, metal thickness, metal roughness, plating thickness, etc. were necessary to drive structure simulators such as HFSS and CST. Despite helpful cooperation from the PWB fabricator, gathering the appropriate data to support this analysis proved problematic for several reasons.

While all of these parameters were measurable, many of the required variables were simply too difficult to measure in sufficiently large quantities and with sufficient accuracy to create reliable SI variation models. For example, line width measurements (critical for impedance and loss calculations) could only be measured by manually placing etched cores under microscopes. This is very slow, interrupting the normal PWB design flow. The typical automated optical inspection (AOI) machine (preferable for measuring large quantities of lines) is targeted at finding gross errors and has a large measurement uncertainty relative to the uncertainty required for accurate impedance estimation. Surface roughness could only be measured with a profilometer, a manual measurement not useful for obtaining statistically significant quantities. Dielectric core thickness is a parameter many fabricators measure only in occasional inspections to verify compliance of incoming materials. Likewise, dielectric permittivity was measured only occasionally on specially constructed PWBs with dielectric test coupons.

Even if large quantities of data were available, some process variables lack statistical normality needed for extrapolation to $5\sigma$ limits. Many factors can potentially cause non-normal characteristics, including sorting etched cores, rejecting cores, or changing dielectric materials.

## SI Variability from Direct SI Measurements

Another approach to analyze PWB variability is to directly measure 4-port S-parameters of every high-speed net across many PWBs and analyze the resulting variability. In addition to the insights offered by S-parameters themselves, other signal integrity variables (calculated from S-parameters) are invaluable in estimating multi-gigabit performance. These include random phase skew, differential-to-common mode conversion ($S_{DC}$), eye diagram metrics (eye height, width, jitter, vertical eye noise), and impedance. The problem, however, is not dreaming up interesting things to measure, but rapidly collecting large quantities of calibrated 4-port S-parameter measurements. Recently, two new tools were developed that enabled acquisition of such large sets of data: a high throughput automated S-parameter test system, and associated system-level compliance test software.

**Automated PWB Test System**
In [2], a large production test system is described, capable of measuring calibrated 4-port S-parameters (800 nets/hour) of active image area high-speed nets on large PWBs (2 feet x 3 feet). The system is equipped with the following key features: Agilent 20 GHz VNA; dual gantry robotic arms; customizable, high-speed pogo-style, all-pin-compliant probe; fully-automated SOLT probe-tip calibration; and compliance testing software (described below), for real-time pass-fail evaluation of each net. While the tester measurement repeatability is quantified in greater detail below, the insertion repeatability is approximately ±0.25 dB from 10 MHz to 6 GHz.

**Compliance Test Software**
The test system described above employs compliance test software to evaluate PWB performance real-time (during test), net by net.  In the simplest case, the software can examine frequency windows of S-parameters or time windows of impedance.  However, as discussed in [3], a rapid link analysis algorithm was developed to simulate eye diagrams in < 1 sec using stored driver waveforms and channel S-parameters.  The test system measures the PWB S-parameters, cascades them with additional S-parameters representing other link components, and simulates a link eye using the driver waveforms.  Eye metrics are quickly computed and the full channel performance is assessed.  In this fashion, nearly all high-speed PWB nets are measured and validated in the context of full link performance.  The intent of this type of production testing is to screen PWBs for high-speed performance anomalies before expensive assembly steps, guarding against manufacturing defects and variability (not detectable with traditional test methods) that impact high-speed performance.

While this software runs real-time within the tester, we employed the software offline to post-process thousands of S-parameter measurements and calculate eye diagrams and the eye diagram metrics.  The repeatability of the system (same net variation across many calibration cycles) in terms of vertical eye opening was approximately ±1% of the nominal eye opening.  Additional detail about the testing and test results is provided in the section, *Analysis of Eye Diagrams*.

## Description of PWBs Analyzed

The PWBs analyzed here were approximately 2 feet by 3 feet, containing over 2000 high speed differential nets per board.   High-speed transmissions line lengths ranged from approximately 1.7 inches to 32.8 inches.  Signal lines were built in an offset stripline stackup as 6 mil wide lines on a 15 mil pitch with 1 oz copper.  In this stackup, signal lines were etched on both sides of a single core and then sandwiched between two pre-preg layers.  This assembly, in turn, was sandwiched between two ground planes, yielding two distinct stripline routing layers sharing common grounds.  Collectively, we refer to two routing layers (built on opposite sides of the same core) as a "routing core", since in our analysis we were able to distinguish nets by routing core, but not by the exact routing layer.  The core and prepreg material both were Nelco 4000-13 ($\varepsilon r \approx 3.4$, $\delta \approx 0.007$ through 2.5 GHz), with a 2113 core weave and a 1080 pre-preg weave. Both weaves are relatively "loose" with respect to the transmission line width and pitch, allowing for potential phase skew effects [16].  The PWB artwork was not rotated relative to the glass weave.

The board consisted of two sub-composites with 4 cores (8 routing layers) each.  The final PWBs were constructed by joining the 2 thick sub-composites with conductive paste using technology discussed in [9].  This minimizes long via stubs with no backdrilling in this case.  Thus far, the circuit connections created by the cured conductive paste have been tested and exhibited no measurable effect using conventional signal integrity test techniques. The finished board had a thickness of approximately 0.2" and each sub-composite was approximately 0.1".  This resulted in worst-case stub lengths ~90 mils, acceptable for the targeted 2.5 Gbps but not acceptable for higher data rates.

## Database of S-parameter Measurements

For our analysis, a total of 11,961 4-port S-parameters were measured across 14 unique production-quality PWBs.  Six PWBs were measured on the automated PWB tester described above for a total of 11,118 measurements (typically 2000 nets per board).  Nine PWBs were tested on a manual test system for a total of 843 nets.  For practical reasons, there are only 85 manual measurements per board, strategically chosen to represent a variety of net lengths on different signal routing layers.  One PWB was tested on both systems.  The manual test system utilized different probes and calibration substrates (custom 40 GHz GGB skate-action probes) that were measured by hand.  The repeatability of this test system was slightly better than the automated tester, as shown in **Figure 2**.

The layout database for this PWB was used to extract descriptive net information such as net length (P and N separately), signal routing core, and net name. Added to this database were PWB serial number and the tester used to measure each S-parameter file.

The automated tester collected data from 10 MHz to 6 GHz while manual data was collected through 10 GHz.  However, all data reported here was analyzed in an identical fashion through 6 GHz only.  We hope that designers can use data reported here to shape realistic expectations about PWB performance variation and accommodate designs to account for it.

## Outcomes and Predictor Variables

Statistical analysis techniques make use of models such as linear regression.  As with any equation model, there are dependent (SI) outcome variables (such as eye height, impedance and phase skew) as well as independent predictor variables (such as net length and board serial number).  Variation of SI outcomes can be thought of as a superposition of random effects and non-random effects.  One important goal of this work is to determine which independent variables cause statistically significant trends in the SI outcome (non-random effects) in order to better understand them, to prioritize them and ultimately control or design around them. In the end, variability in the SI outcomes is either attributed to independent predictor variables or it is left unexplained; the leftover modeling error presumably represents random SI variability.

In this analysis, the outcome variables are tangible signal integrity performance variables with which engineers are familiar.  Outcomes such as frequency-dependent S-parameters and eye diagrams were necessarily reduced to scalar quantities better suited for statistical analysis.  Each eye diagram was sampled to obtain 4 scalar quantities:  eye height, eye width, deterministic jitter and vertical eye noise.  From the S-parameters, the dB/inch differential insertion loss, and differential to common mode conversion ($S_{CD21}$) were all sampled at 1 GHz, 2 GHz and 4 GHz and the scalar dB values were used for statistical analysis.  Random phase skew was also sampled at 1 GHz, 2 GHz and 4 GHz.  Odd mode transmission line impedance (half the differential impedance) was computed in MATLAB from the S-parameter measurements.  All of these quantities were computed for every measured net on every board.

The independent predictor variables considered were as follows:  Net length, routing core, and serial number.  In one case, we also use random skew as a predictor variable for differential to common mode conversion, $S_{CD21}$.

For brevity, we use μ to signify a population mean, σ to signify standard deviation and n to signify the number of measurements or calculated terms.

## SI Variability Analysis Techniques

The measured and computed outcomes described above formed the basis of our SI variability analysis of the PWBs. A multi-variable analysis of variance (n-way ANOVA or "anovan" in MATLAB) was performed for each SI outcome. An ANOVA analysis reveals whether changes in an independent variable (net length, board serial number, signal routing layer, etc.) predict statistically significant changes in an outcome variable (eye height, phase skew, impedance, etc).

More powerful than a simple t-test (which detects statistically significant differences in mean values of a single independent variable), an ANOVA test can operate on multiple continuous variables (e.g. net length) as well as nominal variables (e.g. board serial number). Results of an ANOVA analysis of eye height, for example, can suggest which independent variables cause statistically significant changes in eye height.

Using an ANOVA, a multivariate linear regression model is generated for one SI outcome per analysis (such as eye height) as a function of any number of independent variables (e.g. net length, routing core, board serial number and test system used). A least squares solution provides the best-fit coefficients (β) for the model, as shown below. The model always fits the measured data with some residual error. The residual is simply the error that the model can not predict, given by ε = Measured Outcome – Modeled Outcome.

$$Modeled\ SI\ Outcome = \beta_0 + \beta_1 \cdot \text{var}1 + \beta_2 \cdot \text{var}2 + ... + \beta_N \cdot \text{var}N$$

**Statistical Blocking and Inflated Values of σ**

The modeled SI outcome (e.g. vertical eye opening) may have a linear dependence on any number of independent variables (e.g. net length). Importantly, a large portion of variation may not be random in nature, as shown in **Figure 10** and simple standard deviation calculations often fail to take this into consideration and can not be used for Gaussian extrapolation. Therefore, any linear trend effects from the independent variable can artificially inflate estimates of σ.

Application of ANOVA techniques reduces this problem, and an overall mean-squared error (MSE) of this residual is calculated, which serves as an estimate of the standard deviation (σ) for the outcome variable. This is a key component of the ANOVA, as it indicates the variance in the SI outcome after systematic (linear model) effects have been removed or "blocked" from the data. Effectively, ANOVA models and removes linear trend effects (like the one shown in **Figure 10**) from multiple independent variables simultaneously. This has important implications when using variance to extrapolate 5·σ Gaussian variation. The ANOVA MSE residual (σ) is a much better estimate of random standard deviation than a simple σ estimate computed from global data because non-random linear effects of multiple independent variables (all of which artificially inflate σ) are removed. ANOVA is only one technique for blocking linear effects in variance estimates, and we illustrate a second technique (pooled variance) later.

**F-Ratio**
ANOVA calculates an F-ratio for each independent variable which is simply the factor by which the model's residual MSE increases when that variable is removed from the full model. The F-ratio therefore measures the *relative* importance of independent variables in influencing an SI outcome. An F-ratio of 1 implies that removing the independent variable from the linear model has no effect on the model error, and therefore, the variable has no predictive value. [5] If none of the independent variables are predictive of the SI outcome, it is possible to have a high F ratio for a non-predictive variable, in this case, only suggesting that it is more important than other (non-predictive) variables. ANOVA also calculates a p-value for each independent variable. The p-value indicates whether there are statistically significant changes in the SI outcome due to changes in the independent variable. A p-value of 0.01 indicates there is a 1% probability that the change in the SI outcome happened by random chance alone. Therefore, very low p-values express high confidence that the independent variable (e.g. net length) is significantly related to the SI outcome (e.g. eye height). Thus, important independent linear predictor variables usually have low p values and high F ratios.

**Test System Repeatability**
Test system repeatability is also an important consideration, as some portion of random variation comes from the test system. Wherever we report the variability in an SI outcome, the tester repeatability is also reported. The repeatability of jitter measurements, for example, was computed by calculating jitter from S-parameters measured repeatedly (across many calibration cycles) on a single set of test coupons.

**Distribution Characteristics**
The assumption underlying the linear regression models used in ANOVA is that the data has a normal distribution. As it turns out, most measured/calculated SI outcomes are not strictly normal according to a chi square normality check. This is neither uncommon in large data sets nor a cause for alarm. It has been noted in statistical texts that the conclusions of a linear regression model can be reliable if the assumption is not *badly violated* **[5]**. However, "badly violated" is not well defined, and readers are urged to be cautious when interpreting ANOVA results derived from non-normal populations such as eye width and jitter. Nonetheless, regression models have a certain inherent stability and robustness to them (as do ANOVA conclusions based on regression models) and reliability and accuracy are not entirely diminished, simply because data is not strictly normal by the standards of a normality test [5].

With data sets this large, simple chi-square normality tests almost always fail unless the data is nearly perfect. Therefore, we report more descriptive metrics such as skewness and kurtosis. Skewness measures whether data histograms are heavily weighted to the left or right side of the mean. A skewness of zero indicates no skew. A negative or positive skewness signifies that the data leans heavily toward the left or right of the mean respectively. Skewness is apparent in quantities such as loss-per-inch, since there are physical reasons why loss may increase unbounded but can not decrease below zero.

Kurtosis describes the peak and tails of a distribution in the following way: high kurtosis (> 3.0) signifies a distribution that is more peaked with wide spanning tails that heavily weight the variance with extreme outliers. Low kurtosis (< 3.0) signifies a distribution with a plateau top with typically narrow tails that are spread out less than a normal distribution. A normal distribution has a kurtosis of 3.0. Kurtosis serves as a warning against blindly using N·σ limits from Gaussian statistics for determining worst-case outliers. Data with high kurtosis may provide unpleasant outlier surprises for those who blindly use Gaussian statistics. Several measurements we present having high kurtosis do have outliers exceeding 5σ. This would be a highly improbable event (under normal/Gaussian assumptions) but not surprising in data with high kurtosis.

## Summary of Statistical Analyses

In this section, we summarize our statistical results in tabular form. Three key analyses are summarized in 3 tables below: Same-net variability across boards, global variability of SI outcomes, and finally the ANOVA summary. Rather than discussing all the table results in detail, we selected some of the more interesting cases from the three tables and discuss them in the following analysis sections on S-parameters, phase skew, impedance and eye diagrams.

**Global Variability Summary Statistics**

Below, **Figure 1** summarizes global variation statistics for all 11,960 measurements. Global variability is a raw summary statistic describing gross variation across all data. The min, max, µ and σ are reported as a way of describing the overall data. Global statistics are helpful in comprehending the magnitude of total variation in engineering terms. However, as systematic effects have not been blocked or removed (the impact of length on eye opening, for example) the data has non-random effects built in. Therefore global statistics are the least reliable for assessing purely random variation and extrapolating future performance with 5σ confidence intervals.

| Data Source | EIT Board Measurements | | | | | | |
|---|---|---|---|---|---|---|---|
| Tester Type | Manual + Automated Testers | | | | | | |
| Sample Size | 11,961 Nets / 14 Boards | | | | | | |
| Calculation | Mean | Std Dev | Min | Max | Tester Repeatability (Std Dev) | Skewness | Kurtosis |
| Random Skew (ps): 1 GHz | 3.7 | 12.8 | 0.0 | 56.1 | 2.5 | 0.2 | 3.0 |
| Random Skew (ps): 2 GHz | 4.9 | 12.7 | 0.0 | 56.8 | 1.9 | 0.2 | 3.0 |
| Random Skew (ps): 4 GHz | 4.9 | 12.3 | 0.0 | 60.0 | 2.3 | 0.2 | 3.2 |
| Eye Height (volts) | 0.783 | 0.109 | 0.404 | 1.121 | 0.010 | 0.1 | 2.2 |
| Eye Width (UI) | 0.538 | 0.087 | 0.280 | 0.960 | 0.097 | 1.3 | 7.2 |
| Eye Jitter (UI) | 0.078 | 0.026 | 0.030 | 0.190 | 0.005 | 0.6 | 2.5 |
| Vertical Eye Noise (volts) | 0.153 | 0.045 | 0.026 | 0.260 | 0.005 | -0.2 | 2.2 |
| SDD21 dB/in: 1 GHz | 0.15 | 0.03 | 0.000 | 0.379 | 0.013 | -0.3 | 12.9 |
| SDD21 dB/in: 2 GHz | 0.29 | 0.07 | 0.011 | 0.874 | 0.036 | -3.2 | 19.1 |
| SDD21 dB/in: 4 GHz | 0.42 | 0.12 | 0.003 | 1.714 | 0.077 | -2.4 | 24.7 |
| SCD21 (dB): 1 GHz | -30.4 | 5.3 | -52.1 | -5.6 | 2.7 | -0.2 | 3.1 |
| SCD21 (dB): 2 GHz | -26.4 | 4.6 | -41.7 | -5.6 | 2.3 | 0.0 | 3.2 |
| SCD21 (dB): 3 GHz | -25.3 | 4.4 | -40.7 | -5.6 | 2.1 | 0.0 | 3.1 |
| Impedance (Ω) | 52.4 | 1.5 | 47.5 | 57.6 | 0.24 | 0.0 | 2.3 |

MAYO PROPRIETARY  SEP_28 / 2009 / KJB / 40026v2
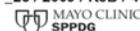

**Figure 1:** Global summary statistics of 11,961 nets across 14 PWBs (40026v2). Impedance statistics came from a smaller sample of 843 nets.

For convenience, the automated tester repeatability (σ) is recorded in a column for comparison to the measured σ of the SI outcomes. For a few SI outcomes, tester repeatability adds significantly (20% or more) and for eye width, the tester repeatability (σ) actually dominated the measured eye width σ. Still, the mean eye opening was 0.538 UI, so the eye width metric was still useful for assessing board performance despite the higher tester variability. The repeatability of the manually collected measurements (n=843) is substantially better, as discussed in the next section.

**Same-Net Variability (SNV) Across Boards**
The SNV is an average measure of how any given net varies across different boards. SNV is useful because (like ANOVA) it removes or blocks systematic effects of variation due to signal routing layer, net length and designed-in features in the layout artwork. Thus, SNV is an excellent measure of variability blocked for most every systematic variation except board serial number. Therefore, it reveals the repeatability of the manufacturing process in building a given structure.

The SNV of each SI metric is based on a widely-known variance ($\sigma^2$) computation called the "pooled variance". First, the outcome variability ($\sigma^2$) across PWBs is computed separately for individual nets on the board, resulting in 11,961 individual variance calculations. Next, all the variances are averaged or "pooled". Not all nets had an equal number of measurements, so individual net variances are weighted by the number of observations per net. Note that individual nets were only measured across 6-14 unique PWB measurements. However, "pooling" thousands of variances significantly boosts statistical confidence because standard deviation estimates (from any sample size) have a centralizing tendency to represent the population standard deviation. In other words, accurate standard deviation estimates can be obtained either by a) calculating a single standard deviation estimate from large sample size or b) averaging a large number of standard deviation estimates, each of which may be based on small sample sizes. Pooled variance uses the latter approach to leverage statistical confidence.

| Data Type | EIT Board Measurements | | | Tester Repeatability | |
|---|---|---|---|---|---|
| Tester Type | Both Testers | Automated Tester | Manual Tester | Automated Tester | Manual Tester |
| Sample Size | 11,961 Nets 14 Boards | 11,118 Nets 6 Boards | 843 Nets 9 Boards | 27 Nets, 10 Trials | 1 Net, 78 Trials |
| Calculation | SNV σ | SNV σ | SNV σ | σ | σ |
| Random Skew (ps) | 7.2 | 7.2 | 5.5 | 2.5 | 0.6 |
| Eye Height (Volts) | 0.024 | 0.024 | 0.004 | 0.010 | 0.003 |
| Eye Width (UI) | 0.069 | 0.069 | 0.037 | 0.097 | 0.011 |
| Eye Jitter (UI) | 0.007 | 0.007 | 0.005 | 0.005 | 0.005 |
| Eye Noise (Volts) | 0.010 | 0.011 | 0.003 | 0.005 | 0.001 |
| SDD21 dB/In, 4 GHz | 0.086 | 0.072 | 0.041 | 0.077 | 0.023 |
| SCD21 (dB) | 4.3 | 4.2 | 3.8 | 2.7 | 1.0 |
| Impedance (Ω) | X | X | 1.1 | X | 0.24 |

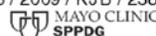

SEP_28 / 2009 / KJB / 23897v2
MAYO CLINIC
SPPDG

**Figure 2: Same Net Variation (SNV) Across Boards (23897)**

We summarize the SNV in **Figure 2**. We report the numbers as σ instead of $σ^2$ to maintain units consistent with the SI outcomes. As an example of how to interpret **Figure 2**, the eye height SNV σ=0.024V while the automated tester repeatability (σ) was 0.010V. We subtract the impact of tester variability by subtracting variances ($σ^2$). Therefore, the true eye height σ is the square root of $(0.024^2 - 0.010^2) = 0.022V$. With a mean eye height of 0.783V (**Figure 1**), this implies a 5σ variation of 14% eye height on precisely the same net (across boards).

Repeating this calculation for the manual tester with a much smaller number of measurements (843 manual versus the 11,118 automated nets), we find the 5σ variation is approximately 2%. Both sets of measurements have tester variation removed, yet the σ estimates, derived from very different numbers of measurements, is substantially different.

**Sample Size versus σ Experiment: How Many Samples are Required?**
Standard deviation (σ) estimates can depend substantially on sample size, as shown above. A standard deviation (σ) estimate based on small numbers of measurements can be misleading, since it is only an *estimate* of the true σ of the entire pool of all measurements. To illustrate this point, we ran a theoretical sampling experiment from a complete set (N=2077) of board measurements. Since the board was fully characterized, the true σ of every SI parameter was known. We compared the known σ to *estimates* of σ based on smaller, random samples of the complete data set. The point was to examine the error in σ *estimates* as a function of the sample size. How many samples are really necessary to get an accurate estimate of PWB variability?

We varied sample sizes from n=10 to 2000. At each value of n, exactly n samples were randomly selected from a pool of 2077 possible measurements and σ was calculated. We repeated this 500 times at each value of n to show the spread of σ estimates at every sample size n. Estimates of σ were centralized around the true population σ but contained spreads with up to ±70% error (n=10) and up to ±35% error (n=30). For n=1000, σ estimates were generally constrained within 5% of the known σ. This result was similar for all SI variables such as impedance and eye diagram metrics. The key point is that confidence in any single estimate of σ from measurements is proportional to the sample size n and the error in any given σ estimate can be substantial, even when n=30 (frequently cited as a minimum statistical sample size). This result holds true for any random Gaussian variables. The take-away message is that large numbers of measurements are required to obtain high-confidence estimates of σ needed for predicting long-term variation in SI parameters such as impedance, vertical eye opening, etc. Low-confidence estimates can be obtained with much smaller sample sizes.

**ANOVA Summary**
**Figure 3** below summarizes the results of the ANOVA analysis. A brief tutorial on ANOVA is provided in an earlier section, *SI Variability Analysis Techniques*. These results assess the causes of systematic (non-random) variation in each SI outcome variable. Independent variables are ranked on a relative scale (F-ratio), indicating which variables have the most influence over the SI outcome variable.

Each column is the result of an independent ANOVA analysis on one SI outcome variable. The results from different columns are unrelated. However, one can compare the F-ratios down a single column to assess the relative importance of independent variables in predicting the SI Outcome. The p-value is also reported.

As an example, consider random skew (1 GHz). ANOVA builds a linear model predicting random skew from the independent variables net length, routing core and serial number. Net length has an F-ratio of 1.4 and a p-value of 0.2297. This indicates that removing net length from the linear model only causes a MSE increase of 1.4. The p-value indicates that there is a 22.97% probability that the observed difference in skew resulting from variations in would happen by random chance alone. Therefore, net length would not be regarded as a statistically significant predictor of 1 GHz phase skew. Routing core and serial number, however, both produced statistically significant shifts in random skew. Serial number, with an F-ratio of 31, was a stronger predictor of random skew than routing core.

| Analysis of Variance Summary | | Random Skew (1 GHz) | Random Skew (4G) | Eye Height | Eye Width | Eye Jitter | Vertical Eye Noise | dB/In (1 GHz) | dB/In (4 GHz) | Freq, SDD11 @ -10 dB | SCD21 (1 GHz) | SCD21 (3 GHz) | Impedance (Ω) |
|---|---|---|---|---|---|---|---|---|---|---|---|---|---|
| Net Length | F-Ratio | 1.4 | 0.5 | 153459 | 151 | 70673 | 146718 | 327 | 193 | 2577 | 0.2 | 2498 | 366 |
| | p-value | 0.2297 | 0.4665 | 0.0000 | 0.0000 | 0.0000 | 0.0000 | 0.0000 | 0.0000 | 0.0000 | 0.6570 | 0.0000 | 0.0000 |
| Routing Core | F-Ratio | 7.0 | 6.9 | 65 | 12 | 36 | 105 | 35 | 169 | 768 | 2.6 | 13 | 4 |
| | p-value | 0.0000 | 0.0000 | 0.0000 | 0.0000 | 0.0000 | 0.0000 | 0.0000 | 0.0000 | 0.0000 | 0.0086 | 0.0000 | 0.0001 |
| Serial Number | F-Ratio | 31 | 37 | 114 | 1 | 30 | 106 | 99 | 204 | 600 | 69 | 152 | 3 |
| | p-value | 0.0000 | 0.0000 | 0.0000 | 0.5173 | 0.0000 | 0.0000 | 0.0000 | 0.0000 | 0.0000 | 0.0000 | 0.0000 | 0.0006 |
| Random Skew, (1 GHz) | F-Ratio | | | | | | | | | 65530 | | | |
| | p-value | | | | | | | | | 0.0000 | | | |

Highly Significant Predictive Terms
Analysis Performed on Manual Data Only (N=843)
Analysis Performed on all Data (N=11,960)

SEP_28 / 2009 / KJB / 40027v2
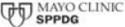
MAYO PROPRIETARY

**Figure 3: Each Column Represents one Unique ANOVA Analysis Summary (40027)**

Below, we present the analyses of key SI outcome variables including S-parameters, phase skew, impedance and eye diagrams. We highlight and expand on some of the more interesting items from the three tables above.

## Analysis of S-Parameters

A variety of quantities can be specified and statistically studied which come directly from the S-parameter data. In this paper, we limit our analysis to a handful of measures which glean information about insertion loss and common to differential mode conversion. Analysis of return loss was more problematic and variable, even when reduced to scalar quantities such as "Frequency at which return loss crosses -10 dB". A full explanation of this problem is outside the scope of this paper.

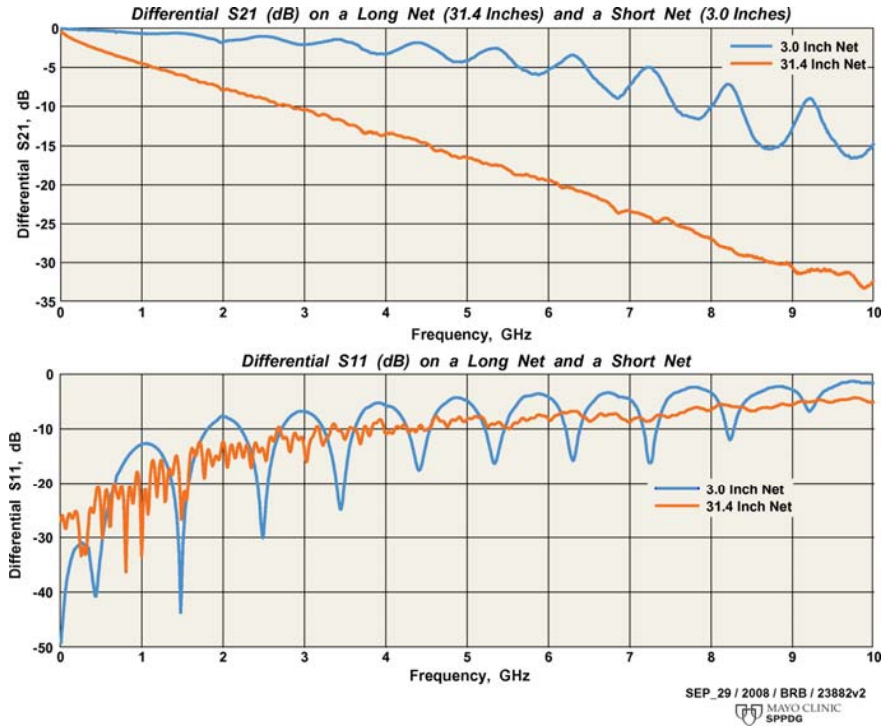

**Figure 4: SDD21 and SDD11 Illustrating the Basic S-parameter Characteristics for a Long and Short Net (23882v2)**

Direct S-parameter analysis will be related to other computed parameters such as eye height, eye noise, and phase skew. However, it is hoped that direct statistics on the raw data will shed additional light on the perhaps subtle manufacturing variations which affect the electrical measurements. **Figure 4** shows plots of fairly typical $S_{DD21}$ and $S_{DD11}$ measured on the PWBs we examined. The overwhelming majority of the S-parameter data was collected using the large automated test system [2]. The large vias in this stackup result in capacitive reflections and therefore interference patterns (constructive and destructive) that depend on the wavelength (frequency) and net length. Longer nets suffer the same problem, but transmission-line attenuation significantly reduces resulting interference patterns.

One interesting S-parameter we analyzed was the differential to common mode insertion conversion, $S_{CD21}$. We examined the impact of phase skew on $S_{CD21}$ by performing an ANOVA analysis. In this analysis, phase skew is regarded as an independent parameter and $S_{CD21}$ is the SI outcome. As discussed above, phase skew increases common mode signals on a transmission line. The independent variables considered were net length, routing core, serial number and random phase skew. The ANOVA created a linear model of $S_{CD21}$ in terms of the independent variables (length, serial number, etc) and the MSE between the model and the measured data (i.e. the variance) was computed.

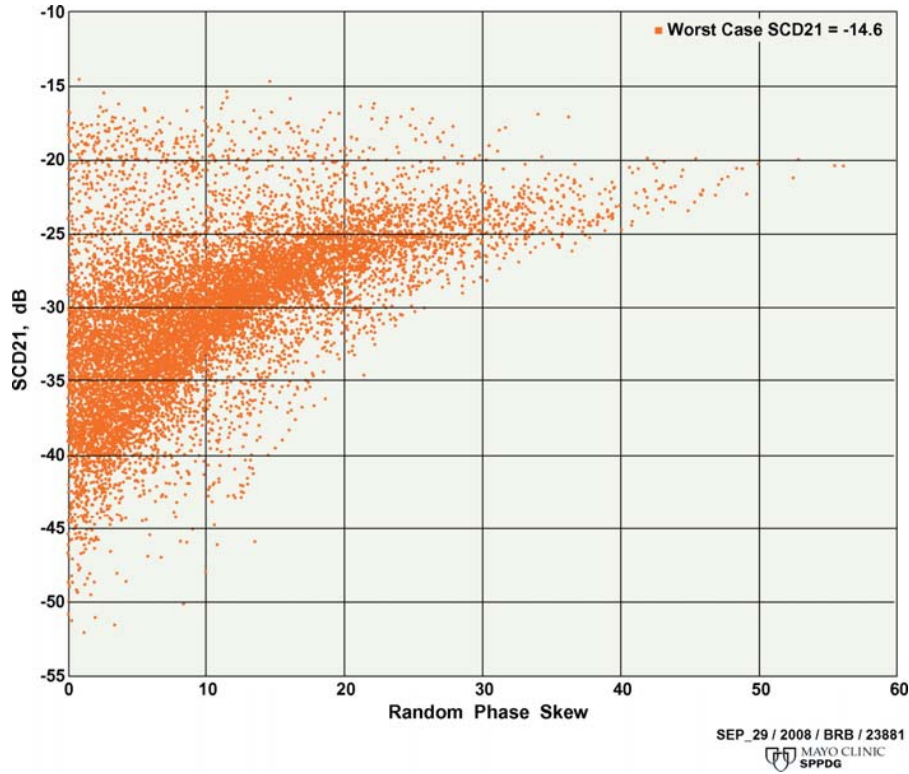

**Figure 5: Common Mode Conversion Versus Random Phase Skew (23881)**

Since the two quantities are related mathematically, we expect some correlation. The ANOVA indicated that phase skew had an F-ratio of 65,530, which indicates that the MSE in the linear model ($S_{CD21}$ in terms of phase skew, routing core, etc) increased by a factor of 65,530 when phase skew was removed as a predictor variable. The F-ratios of net length, routing core and board serial number were 0.2, 2.6, and 69 respectively. These statistics confirm the strong association (predictive correlation) between phase skew and $S_{CD21}$. The scatter plot of $S_{CD21}$ vs random phase skew in **Figure 5** further elucidates the relationship in a way the statistics (and the direct mathematical relationship) do not reveal.

The low phase-skew portion of the chart implies that low phase skew is a necessary, but not a sufficient condition for low $S_{CD21}$. Other (unknown) factors also contribute to high $S_{CD21}$, even in low-skew conditions. The high phase-skew region of the chart implies low phase skew is no guarantee of good conversion, but high phase skew assures the worst-possible $S_{CD21}$.

## Analysis of Phase Skew

Another quantity of interest is phase skew, which is the difference in transit time between the P and N nets. Phase skew has two deleterious impacts on signal integrity performance of digital systems: reduced horizontal eye opening and increased differential to common-mode conversion, which exacerbates crosstalk. Both are the result of one signal edge rising before the complementary edge begins to fall.

**How It's Calculated**

Total time skew is computed from measured 4-port S-parameters as the difference in time of flight in the P and N conductors. The P conductor time of flight is calculated from $S_{21}$ as unwrapped_phase($S_{21}$)/360·Period, where Period is the reciprocal of the frequency. The N conductor flight time is similarly calculated from $S_{43}$. The total skew may be positive or negative and it is measured in units of picoseconds (ps).

Differences in P and N physical net lengths (due to routing limitations) contribute to a "designed-in" skew value during layout. We account for this skew and we compute this skew using the known P-N length mismatch (from the PWB layout database) divided by the average propagation velocity of the longest net. The maximum designed-in skew was 8 ps.

Skew resulting from differences in dielectric environment is more subtle, but contributes the most variation and uncertainty in total skew. We refer to this as "random skew", recognizing it may contain non-random components. It is calculated using the relation:

> Random skew = Total skew - Designed-in skew

The automated tester data (n=11,118 nets) showed a small bias of approximately +5 ps random skew while the manual test data (n=843 nets) showed a bias of –4 ps random skew.

**Origin of Phase Skew**
Random phase skew is believed to arise from the glass fiber weave in fiberglass dielectrics. As glass and resin polymers have a very different dielectric constants, localized weave variations cause localized permittivity (and therefore propagation velocity) variations, as addressed previously [4, 10, 12, 13, 14]. If the P conductor should align over a glass fiber while the N conductor is over pure polymer, then the P and the N lines will have different propagation velocities. Since the glass/resin permittivity ratio could easily be 2.0 or more, propagation velocities could (in theory) be √2 times different in P and N complements.

Thought experiments about long lines lying parallel to weave have led to some conventional wisdom that phase skew should increase with net length. In theory, this could potentially result in skew accumulation of 55 ps per inch of line. Experimental evidence was presented in [10] suggesting that the eye height impact of fiberweave increased proportionally to net length while eye width impact of fiberweave increased exponentially with trace length.

**Phase Skew Data Analysis**
The σ of the global phase skew data (**Figure 1**) was 12.8 ps (12.6 ps after removing tester variability). The statistically blocked same-net variability (SNV) across boards (**Figure 2**) shows a σ=7.2 ps (6.8 ps after removing tester variability). Only the σ=6.8 ps number can be used to estimate 5σ confidence intervals (5σ=34 ps) since global data includes unblocked non-random effects. The ANOVA results in **Figure 3** do indicate that routing layer and board serial number had important non-random effects (explaining the wider global variation) that widen the overall variation. Indeed, the global min/max data of **Figure 1** shows measured skew numbers of of 55-60 ps, similar to the high skew results reported in [11]. Again, this PWB artwork was not rotated.

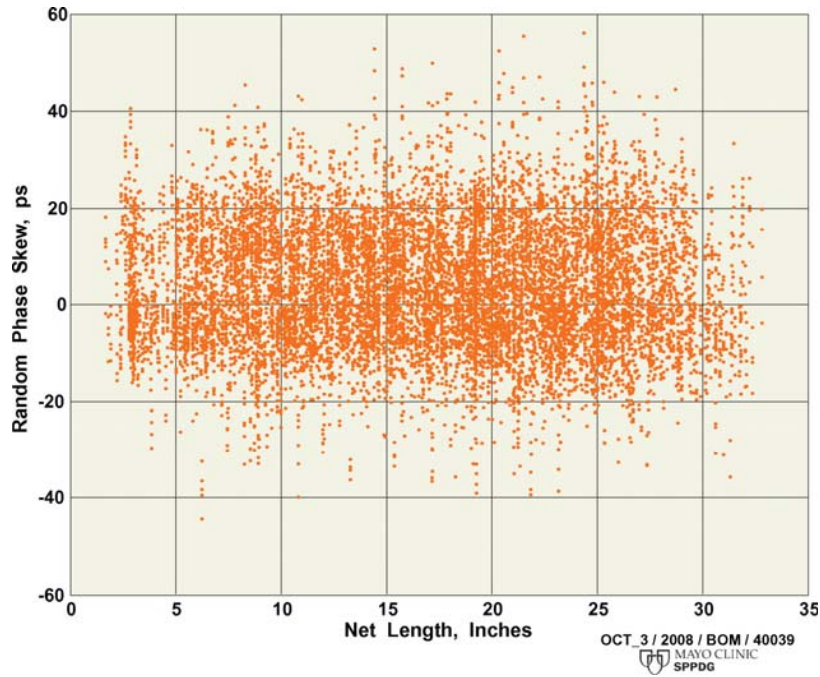

**Figure 6: Phase Skew (sampled at 1 GHz) vs Net Length, 11,960 Measurements (40039)**

Such high skew numbers make sense in light of the distribution which had very long, Gaussian-like tails, reflecting the diminishing (but very real) probability of observing very large phase skews in boards containing large numbers of nets.

**Phase Skew Versus Net Length**
In our ANOVA analysis, net length (ranging from 1.7" - 32.8") was not a predictor of phase skew (p=0.4665, F=0.5). Of all the SI outcomes analyzed, phase skew was the only SI variable not showing significant dependence on net length. A scatter plot of 11,960 measurements in **Figure 6** visually confirms the ANOVA skew results in **Figure 3**. Weak correlations between phase skew and net length were reported earlier in [4], based on smaller samples (n=550, 6 boards) of the same data set analyzed in this paper. Upon further analysis of larger sample sizes (n=11,960, 14 boards), statistical indicators such as the F-ratio and p-value in **Figure 3** suggest that such a correlation is less certain for our data set.

As a further investigation, we were able to study the dependence of phase skew variation on net orientation (routing angle). Two of the 8 routing cores on this board contained approximately half of the nets routed on 45º diagonals. If weave effect dominated random phase skew (assuming 45º diagonal traces should reduce skew), then one would expect these routing cores to show less skew variation. However, the ANOVA analysis indicated only a weak (but statistically significant) dependence upon routing core. **Figure 7**. depicts the phase skew variation on 6 boards (11,118 measurements), stratified by routing core. Half the nets on cores 1 and 3 (orange) were diagonal routes. Diagonal cores 1 and 3 do not appear to have smaller σ. To the contrary, core 3 across 6 PWBs had some of the largest standard deviations among all cores while core 1 appears in line with other cores.

The fact that phase skew had no correlation to net length and the fact that diagonal routes did not reduce the overall skew variation are both puzzling in light of conventional wisdom about fiber-weave skew. The data suggest that small nets can contain as much random phase skew as large nets. This is not unreasonable since P-N phase velocity imbalances (resulting from one complement centered predominantly over resin or glass) do not require large lengths to accumulate phase skews of these magnitudes. The data may simply imply that the probability of this occurrence is independent of net length.

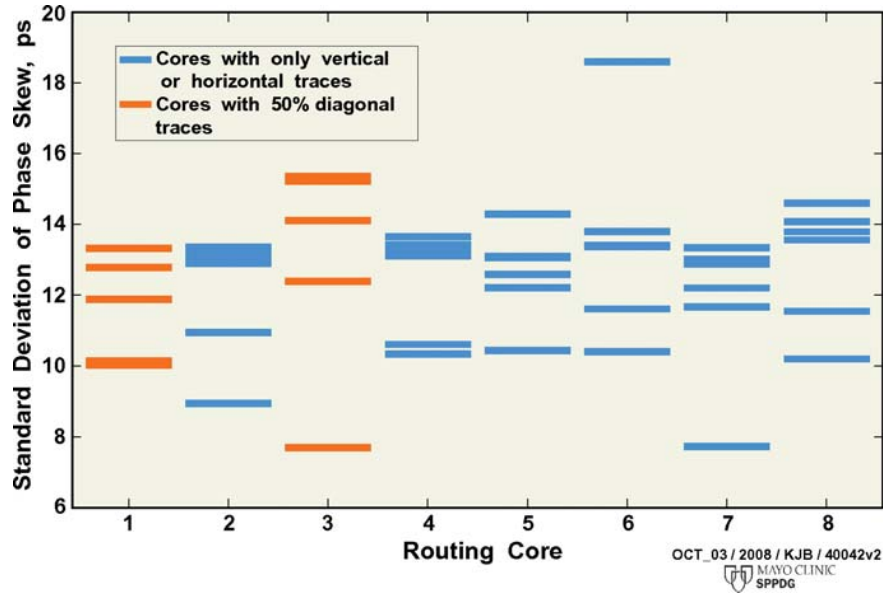

**Figure 7: Phase Skew σ by Routing Core for 6 Boards. Cores 1 and 3 (Orange) Have Half their Nets Routed Diagonally (40042)**

However, the observation that cores with half diagonal routing contain as much phase skew variation as non-diagonal cores remains puzzling. In theory, it is possible that other non-weave effects could impact phase skew. Such asymmetries may arise from asymmetrical via loading from the dielectric, particularly since the 1080 and 2113 glass yarn used has a weave pitch ~16-21 mils, similar to the via diameters. Other types of via loading have been suggested in [4] including asymmetrical routing entry to vias and asymmetrical ground via loading, although experimental evidence in the same paper suggested only sub-picosecond via-induced skew. Nonetheless, dielectric variations around and between vias reported in [8] could explain some random skew. At this time, however, the conclusions are uncertain. More studies of non-weave skew effects would be useful, as artwork rotation is increasingly relied upon to minimize skew for high data rate applications.

## Analysis of Impedance

Impedance is a frequently discussed SI outcome that impacts PWB performance at high frequency. Most high frequency transmission lines are designed with an impedance target which minimizes reflections where the nets meet the driver or receiver. Unlike other SI outcomes, impedance is frequently measured during PWB production, either on cut-away coupons at the board periphery or (less often) in the image area (active circuit area) with flying probe TDR testers. Impedance mismatches in a link result in less energy being transmitted along the intended signal path, a well-known phenomenon called mismatch loss. In the S-parameter frequency domain, reflections are manifest as periodic ripple in both $S_{11}$ and $S_{21}$.

As characterization of transmission line impedance is currently one of only ways to track fabricator performance against SI criteria, transmission line impedance is often measured on periphery coupons, with specifications of ±10%. However, there are important questions to investigate: How does coupon impedance compare to image area impedance? How does coupon impedance variation track image area impedance variation? After all, image area lines are generally surrounded by higher copper density variations that cause etching variation. Therefore, we questioned whether a 10% coupon specification has a meaningful relationship to the variation of image area nets and we discuss it below.

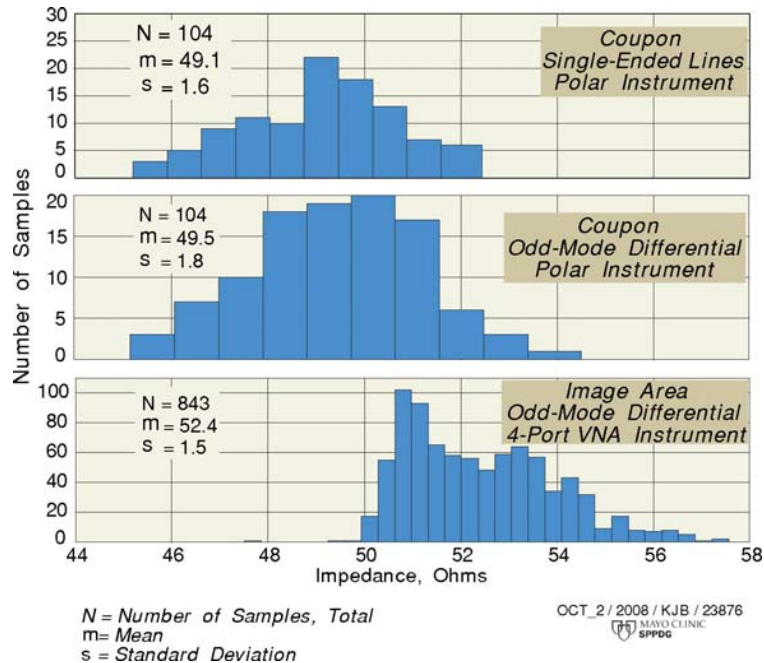

**Figure 8: Coupon versus Image Area Impedance Measurements (23876)**

To help answer these questions, we computed the impedance of 843 image area transmission lines across 9 PWBs. We also collected 104 impedance measurements from cut-out periphery coupons built during the same time period on the same manufacturing line. These measurements are shown in **Figure 8**. The computed odd mode impedance has a mean 2.9Ω higher than the coupons, a statistically significant difference. This may also have engineering significance. However, σ in coupon measurements (σ =1.8 Ω) and image-area measurements (σ =1.5Ω) are very comparable. However, a visual inspection of hundreds of TDR curves showed small impedance perturbations of a few ohms as transmission lines changed routing layers (on buried vias) and traversed through dense pin fields. This suggests that more complex impedance variation effects are possible, but not captured by small time-windowed impedance measurements.

As shown in the section ***Summary of Statistical Analyses***, the SNV in **Figure 2** was σ=1.1Ω, implying a ±11% variation at the 5σ limits. Either a 4σ or 5σ limit would be appropriate, but we elected to use a 5σ limit since our board contained > 2000 nets and a 4-sigma compliance (99.99683%) permits roughly 1 failure for every 31,000 nets. Note that it is not appropriate to use the global σ=1.5Ω from **Figure 1** to extrapolate the 5σ limits, as blocking is not used in the global statistics. Such an extrapolation is inflated with non-random (linear) factors due to board-board differences, net-length differences and routing-core differences, as shown in **Figure 3**. While the global variability is very real, the correct way to consider global impedance variations is to superimpose the 5σ ANOVA model residual MSE (blocked σ) on top of the ANOVA's multivariate linear ANOVA model.

**Measurement Equipment**
Impedance measurements from a PWB fabricator are usually performed using Polar hand-held impedance probes. The coupons measured for this study were measured in this way. These probes perform a time-domain reflectometry (TDR,differential and single ended), where the reflected voltage is mathematically converted to a normalized reflection waveform which can be used to calculate impedance. The waveform is processed by Polar software so as to measure a small time-windowed average impedance at a point corresponding to the center of the transmission line.

The image area nets used in this study were measured using the manual methods described above, using a 20 GHz Agilent 4-port VNA. We computed impedance using the MATLAB (R2007b) RF Toolbox. $S_{DD11}$ was converted to a rational function approximation and the transient step response was calculated to obtain the TDR waveform. Several computed TDR waveforms were validated against TDR waveforms computed in Agilent's Physical Layer Test System (PLTS) software, used for 4-port VNA acquisition and S-parameter post-processing. We computed impedance from a 2 ns time-window average as early as possible after the capacitive effects of the first via settled.

Since raw TDR waveforms exhibit slowly rising impedance (a measurement artifact due to signal loss) the timing of the TDR sample window can impact the computed impedance. One would expect to see higher Polar measurements (sampled in the center of the line) and lower S-parameter computed impedances (sampled near the beginning of the line). However, when comparing the measured coupons (measured with the Polar instrument) to the measured image area nets (measured with the VNA) the opposite is true, suggesting the real difference between coupons and image area nets may be larger than the observed 2.9Ω.

Differences in measurement equipment (Polar hand probes versus VNA S-Parameters) were investigated to see if it explains the 3 Ω difference between coupon and image area impedance. A polar probe hand tester was used to test 170 image-area nets on one PWB's measured S-parameters. The Polar equipment was adjusted to sample the impedance consistent with the way the MATLAB measured impedance from S-parameters (in a small window immediately after the TDR waveform settles). From the Polar measurements, we found μ=52.6Ω and σ=2.4 Ω (n=170 measurements). From the MATLAB- computed impedances from S-parameters, we found μ=52.4Ω and σ=1.5Ω (n=843 measurements). We could not make a direct, net-net comparison in this case. However, we found no statistically significant differences in mean impedances due to equipment types. This implies that equipment differences likely do not account for the 3Ω difference observed between coupon and image area impedance described above.

In conclusion, there may be important differences in impedance between coupons and image area nets, on the order of 3Ω in the case of the boards measured in this work. That difference is not likely explained by test equipment or test method. Both the measured random variation in image-area impedance and the extrapolated 5σ SNV was approximately 10% around the mean with potentially wider global variations.

## Analysis of Eye Diagrams

The eye diagram analysis represents the variability expected in full link performance, based on variable PWB measurements, but with otherwise fixed model components. It is a composite measure of both PWB variability and SI performance. Eye opening and jitter relate directly to system design constraints, so often they represent the bottom line in link performance.

A laboratory measurement of eye diagrams often involves injecting long bit sequences into interconnect fabric, such as the PWB net, and recording the far end waveforms with an oscilloscope. This technique has the attractive feature in that it performs time domain measurements of the actual PWB net. It also has several shortcomings. It only measures the PWB net performance, and while that is important, it omits important channel components, such as connectors and chip packages. It also utilizes a synthetically generated bit stream waveform, which may not match the characteristics of the actual drivers and receivers used in the system.

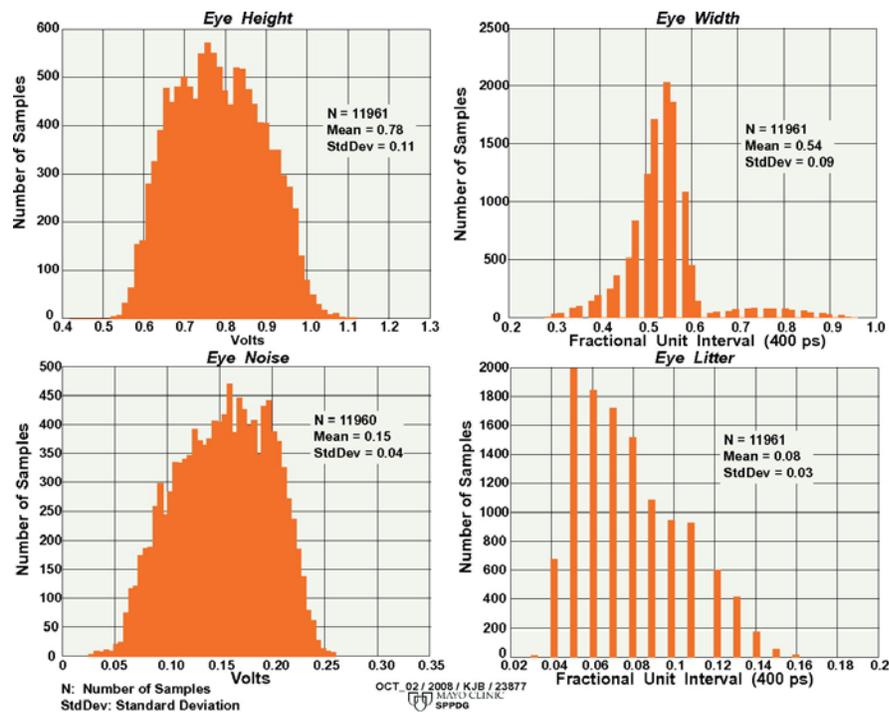

**Figure 9: Histograms of Link Performance Variability in Terms of Eye Diagram Metrics (23877)**

As described above and in [3], the link analysis software used in this study utilizes models of all link components, offering significant improvements over traditional laboratory methods.

**Figure 9** shows histograms of all 4 eye diagram metrics based on 11,960 4-port S-parameter measurements across 14 PWBs. While eye height and vertical eye noise more closely follow a normal distribution, eye width and jitter are highly skewed waveforms.  The discrete appearance of histogram bins of these time-based parameters is a result of both the discrete time nature of the data and histogram binning process.

An ANOVA analysis of eye height, eye width, jitter and vertical eye noise all revealed that net length is easily the most predictive independent variable for all four quantities (F-ratio = 153,459).  The association of eye height and net length is an expected one and a scatter plot showing a clear trend is shown in **Figure 10** below.  However, even here, everything is not as it appears.  Some portion of the variation around a best-fit trend line in **Figure 10** is not random.

Insertion loss of short nets exhibited non-random effects including periodic variation with frequency (ripple in $S_{11}$ and $S_{21}$) as well as periodic variation with net length (observable in **Figure 10**).  This phenomenon is caused by a situation common in many PWBs:  large through-vias (100-200 mil long in this PWB) are attached to opposite ends of the transmission lines.  On small nets (less than 5-7 inches), reflected energy sloshes around with little attenuation, creating interference peaks and nulls that vary with frequency and net length.   Consequently, some small nets (~5 inches) with destructive interference patterns from reflections exhibited eye opening similar to nets 20-23 inchs.   Nets slightly longer (say, 7 inches), having a constructive reflection interference pattern, exhibited eye openings nearly 150-200 mV larger than 5 inch nets, contrary to common intuition.

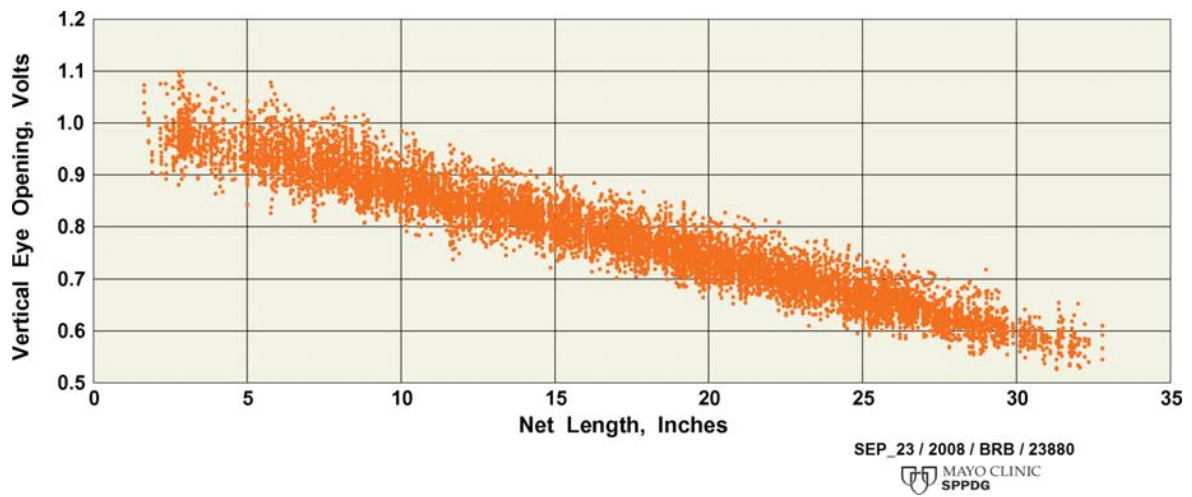

**Figure 10:  Eye Height versus Net Length for 11,960 Net Measurements (23880)**

Length had an F-ratio of 151 for the eye width ANOVA, indicating length was also very significant in predicting eye width.  A scatter plot of eye width versus length (not shown) using the more reliable manual data (n=843 measurements) demonstrate that this association is not a clear, linear trending relationship.   For nets > 12", eye width trends slightly higher with length, moving from approximately 0.5 UI at 12 inches to 0.55 UI at 32 inches.  The variability spans approximately 0.1 UI.  For nets < 12", the variability of the eye width increases substantially, spanning approximately 0.3 UI.  This wider variability at lower net lengths is attributed to the reflective effects discussed above.

Jitter, on the other hand exhibited a very clear linear increasing trend with net length (F-ratio = 70,673), much like vertical eye opening. The same clearly defined trend also characterizes vertical eye noise increasing with net length (F-ratio = 146,718).

The ANOVA analysis also revealed that vertical eye opening had a statistically significant shift in means across board serial number (F-ratio = 114). This is particularly interesting since it suggests shifts in a fundamental SI outcome across different boards in a fabricated lot. A set of histograms in **Figure 11** below plots the vertical eye opening for 6 different boards. Each board (except one) has approximately 2000 eye diagram computations from 2000 independent nets.

In this case, while the results are statistically significant and can be observed in **Figure 11**, a glance at the worst-case performance and spreads do not suggest serious engineering significance. This is a case where we differentiate "statistical significance" from "engineering significance", as they do not always agree.

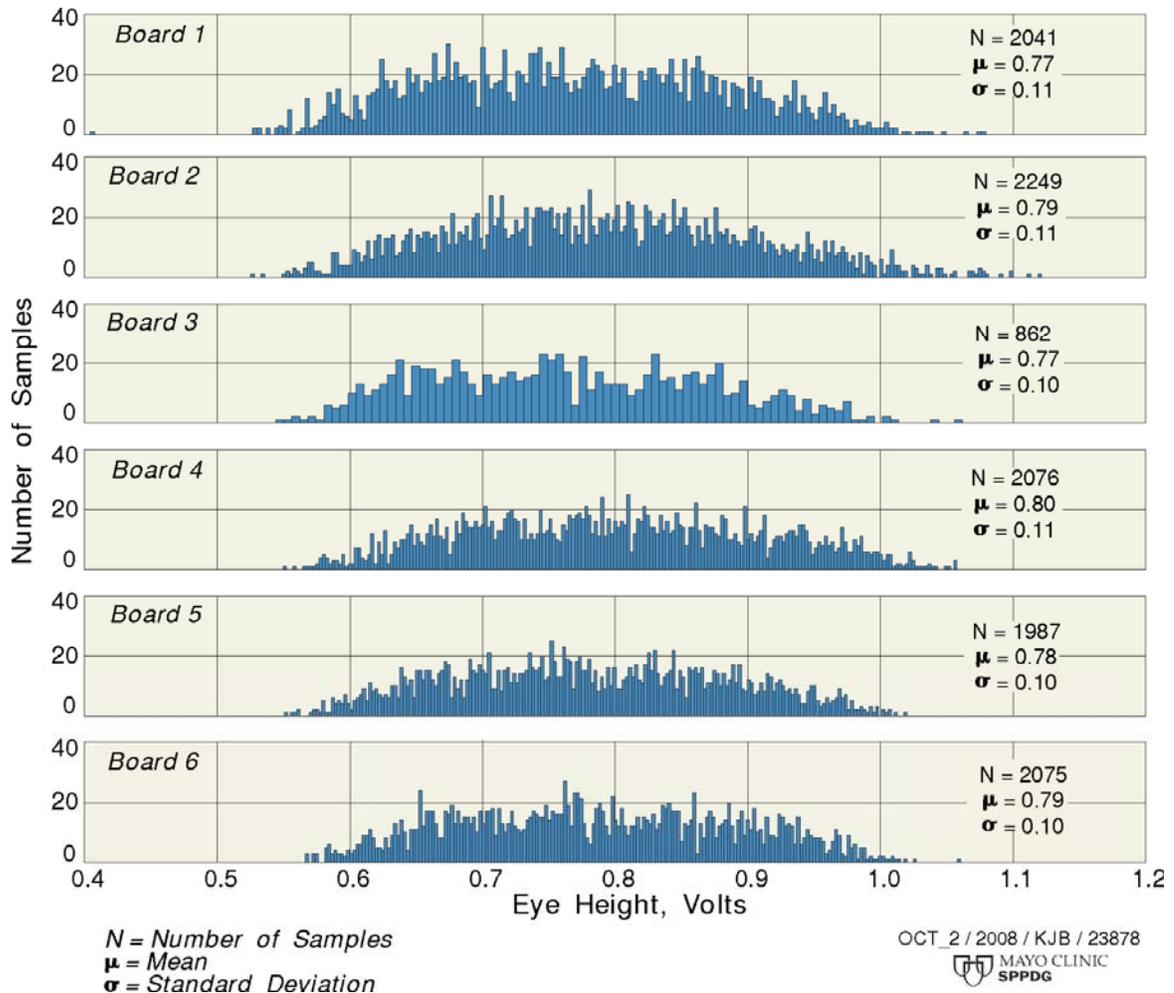

**Figure 11: Histograms of Vertical Eye Openings for 6 Unique PWBs. (23878)**

## Applications of PWB Variability Analysis

A direct analysis of PWB SI variability which we have described above has many applications. As stated in the introduction, it is important for the design engineer to have a good empirical grasp of the real PWB variability to maximize performance from the lowest cost materials, without undue risk.

Link simulations can benefit substantially from the results of direct PWB SI analysis. Simulating $5\sigma$ corner cases for line impedance data is helpful in predicting reflection problems arising from worst-case transmission line impedance. Simulating known $5\sigma$ or worst-case phase skew data simulations would help estimate worst-case crosstalk in sensitive components such as connectors. In both cases, empirically-derived statistics can lend confidence to simulation analyses and help make appropriate choices of board dielectrics, glass weaves and artwork rotation.

One question which vexes engineers is whether combining worst-case corners creates overly pessimistic simulation conditions that would never realistically happen. Using this type of analysis, one can perform correlations or linear modeling between various outcomes in addition to the ones discussed here. When two SI outcomes are uncorrelated, stacking worst-case conditions may not be warranted since the joint probability of both (independent) worst-case occurrences is therefore very low. On the other hand, if (for example) $S_{DD21}$ (db/in) and impedance are correlated and their joint occurance is likely, then worst-case corners may have to be stacked. When SI metrics are causally associated, as indicated by the F-ratio in the ANOVA analysis, then simulating one corner (e.g. high phase skew) may thereby generate the worst-effects of the second corner (e.g. high $S_{CD21}$).

Beyond simulations, hardware test vehicles may also be tested with designed-in worst-case variation. In this way, links can be tested in limited number of corner-case conditions. Direct SI variability analysis, particularly the ANOVA calculation, offers keen insights into what variables have the most impact on design and testing. If eye width is deemed to be a limiting performance outcome, testing may be needed to validate models emphasizing this outcome. But how much testing is needed and how should it be prioritized with limited board space and limited test time? Over what independent variables should one test to best validate eye width? If PWBs are well understood through direct SI measurements, an ANOVA analysis of eye width may suggest, as it did in our case, that net length is the key variable over which eye width should be tested. From the data presented here, scatter plots of eye width versus length further validated that links < 12" inches have the largest eye width variability while links > 12" will likely exhibit the greatest horizontal eye closing.

If S-parameters of all high speed PWB nets are measured and link simulations are run on every PWB net, the direct study of SI distributions may prove valuable for process improvement and PWB screening. The statistical techniques discussed here help identify which SI outcomes are prone to extreme outliers. This type of analysis helps determine the parameters best suited for pass-fail screening of PWBs with a minimal quantity of testing.

## Conclusions

We have demonstrated that large statistical data sets can be acquired on real (not just "test") PWBs using new tools (tester and rapid link analysis software) described above. These statistical data sets can be mined to determine relationship between independent variables and dependent variables critical to SI design and performance.

Several key findings highlight the benefits of analyzing all nets on a board. From the basic S-parameters, we found that to achieve low $S_{CD21}$, keeping low phase skew is essential but not sufficient as other unknown factors contributed to $S_{CD21}$ as well. On the other hand, having higher phase skew virtually guarantees high $S_{CD21}$. Further analyzing all phase skew measurements, we discovered no association between net length and random phase skew. We showed that routing cores containing substantial numbers of diagonal routes displayed as much skew variation as cores containing horizontal/vertical routing only. Together, these observations suggest that there could potentially be additional (yet unknown) contributors of random phase skew other than fiber weave skew. In the worst case, phase skews of 50-60 ps were observed.

An analysis of impedance of 843 nets for same-net variation across boards showed that impedances were within a ±10% specification, although systematic effects of net length will result in global impedance variation larger than that. A comparison of image-area and coupon impedance measurements taken from sets of boards from the same fabricated lot showed that image area nets had an average impedance 3Ω higher than the impedance coupons. We verified that this offset could not be explained by TDR sampling methods or differences in measurement equipment.

By performing ANOVA analyses on different SI outcomes including phase skew, eye diagram metrics, S-parameters and impedance, we used F-ratios to identify key variables that contribute the most variation to the SI outcomes. While eye height, vertical eye noise and jitter had very strong associations with net length, eye width experienced a very nonlinear relationship with net length, exhibiting higher variation of eye width in nets < 12" but with overall worse eye width for nets > 12". Nearly all variables exhibited somewhat weaker, but statistically significant differences in across routing layer and across boards. However, a comparison of eye height histograms revealed that such statistically significant results do not always translate to serious engineering significance.

As data rates reach ever higher, direct and exhaustive SI analyses of PWBs can help design engineers maximize the performance of low-cost materials at reduced risk of failure. Simulation accuracy can be improved while simultaneously avoiding overly-pessimistic estimates of PWB variability. Time and effort for building and testing hardware vehicles can also be reduced while simultaneously increasing confidence that worst-case variability has been analyzed.

## Acknowledgements

The authors would like to acknowledge several people who contributed to the paper in important ways. Mike Degerstrom of Mayo offered many stimulating discussions about the nature of phase skew and its relationship to net length. We thank Kimberly McCoy of the Richard L. Roudebush VA Medical Center and Dr. Jason Sutherland of Dartmouth Medical School for their instructive lessons and suggestions about the statistical analyses. Dan Schraufnagel of Mayo offered many insights about the automated tester and the database of PWB measurements and Devon Post of Mayo offered valuable discussion about PWB construction. We are very appreciative of the Mayo graphic artists Steve Richardson, Elaine Doherty, Terri Funk and Deanna Jensen and we thank them for preparing the outstanding artwork.## References

[1] T. Swirbel, A. Naujoks, M. Watkins, *Electrical Design and Simulation of High Density Printed Circuit Boards*, IEEE Transactions on Advanced Packaging, vol. 22, no. 3, August 1999.
[2] D. Schraufnagel, B. McCoy, W. Fjerstad, D. Johns, B. Gilbert, E. Daniel, *Fully Automated Large Form Factor (2'x3') Four Port Differential 20 GHz Vector Network Analyzer Test System with Real Time Link Characterization*, AutotestCon, 2007.
[3] B. McCoy, J. Coker, B. Techentin, B. Gilbert, E. Daniel, *A Rapid Link Analysis Technique Using 4-Port Scattering Parameters*, 16th IEEE Conference on Electrical Performance of Electronic Packaging, Atlanta, October 2007.
[4] M. Degerstrom, B. Buhrow, B. McCoy, P. Zabinski, B. Gilbert, E. Daniel, *System Level Approach for Assessing and Mitigating Differential Skew for 10+ Gbps SerDes Applications*, Electronic Components and Technology Conference, 2008. ECTC 2008. 58th 27-30 May 2008 Pages 513-520]
[5] Kleinbaum, Kupper, Muller, *Applied Regression Analysis and Other Multivariable Methods*, 2$^{nd}$ Edition, Duxbury Press, 1988, pp.124-125.
[6] R. Hogg, J. Ledolter, *Applied Statistics for Engineers and Physical Scientists*, 2$^{nd}$ Edition, Macmillan Publishing Co, 1992.
[7] MATLAB Help File, version R2007b.
[8] L. Simonovich, *Relative Permittivity Variation Surrounding PCB Via Structures*, SPI2008, Avignon, France, May 12-15, 2008.
[9] R. Das, F. Egitto, V. Markovich, *Nano- and Micro-filled Conducting Adhesives for Z-axis Interconnections: New Direction for High-Speed, High-Density, Organic Microelectronics Packaging*, Circuit World, Vol. 34, No, 1, 2008, pp. 3-12.
[10]. J. Loyer, R. Kunze, X.Ye, *Fiberweave Effect: Practical Impact Analysis and Mitigation Strategies*, DesignCon 2007.
[11]. Heck, H., *et al*, *Modeling and Mitigating AC Common Mode Conversion in Multi-Gb/s Differential Printed Circuit Boards*, Electrical Performance of Electronic Packaging, Oct. 2004, pp. 29-32.
[12]. S. McMorrow, C. Heard, *The Impact of PCB Laminate Weave on the Electrical Performance of Differential Signaling at Multi-Gigabit Data Rates*, DesignCon *2005*.
[13] G. Brist, B. Horine, G. Long, *Woven Glass Reinforcement Patterns*, Printed Circuit Design and Manufacture, November, 2004, pp. 22-33.
[14] Heck, H., *et al*, *Impact of FR4 Dielectric Non-Uniformity on the Performance of Multi-Gb/s Differential Signals*, Electrical Performance of Electronic Packaging, Oct. 2003, pp. 243-246
[15] K. Bois, J. Diepenbrock, J. Loyer, F. Gisin, S. McMorrow, I. Novak, *Panel Discussion: High Speed PCB Traces: What is their True Performance?*, DesignCon 2008.
[16] G. Brist, B. Horineds, G. Long, *High Speed Interconnects: the Impact of Spatial Electrical Properties of PCB Due to Woven Glass Reinforcement Patterns*, IPC Printed Circuits Expo, SMEMA Council APEX, Designer's Summit 2004.
[17] F. Liu, A Practical Method to Estimate Interconnect Responses to Variabilities, Design, Automation and Test in Europe, 2006. DATE '06. Proceedings Volume 1, 6-10 March 2006 Page(s):2 pp.
[18] M. Orshansky, C. Spanos, Hu Chenming, Circuit Performance Variability Decomposition, Statistical Metrology, 1999. IWSM. 1999 4th International Workshop on, 12 June 1999 Page(s):10 – 13.**25 of 25**